\documentclass[letter]{aa}  

\usepackage{graphicx}
\usepackage{txfonts}
\newcommand{\kms}{${\rm\,km\,s^{-1}}$} % kilometres per second

\newcommand{\lppr}{\stackrel{<}{\scriptstyle \sim}}
\newcommand{\lappr}{\raisebox{-0.4ex}{$\lppr$}}

\begin{document} 
\title{Gaia DR2 white dwarfs in the Hercules stream}

\author{Santiago Torres\inst{1,2}\thanks{Email;
    santiago.torres@upc.edu}\and
        Carles Cantero\inst{1}\and
        Mar\'\i a E. Camisassa\inst{3,4}\and
        Teresa Antoja\inst{5}\and
        Alberto Rebassa--Mansergas\inst{1,2}\and
        Leandro G. Althaus\inst{3,4} \and 
        Thomas Thelemaque\inst{6}\and
        H\'ector C\'anovas\inst{7}}
        
\institute{Departament de F\'\i sica, 
           Universitat Polit\`ecnica de Catalunya, 
           c/Esteve Terrades 5, 
           08860 Castelldefels, 
           Spain
           \and
           Institute for Space Studies of Catalonia, 
           c/Gran Capit\`a 2--4, 
           Edif. Nexus 104, 
           08034 Barcelona, 
           Spain
           \and
           Facultad de Ciencias Astron\'omicas y Geof\'isicas, 
           Universidad Nacional de La Plata,
           Paseo del Bosque s/n, 
           1900 La Plata, 
           Argentina
           \and
           Instituto de Astrof\'isica de La Plata, UNLP-CONICET,
           Paseo del Bosque s/n, 
           1900 La Plata, 
           Argentina
           \and
           Institut de Ci\`encies del Cosmos, Universitat de Barcelona (IEEC-UB), Mart\'i i Franqu\`es 1, 08028 Barcelona, Spain
           \and
          Industrial and Informatic Systems Deparment,
            EPF - Ecole d'Ingénieurs, 21 boulevard Berthelot, 34000 Montpellier,
            France
            \and
            European Space Astronomy Centre (ESA/ESAC), Operations Deparment, Villanueva de la Cañada  E-28692 (Madrid), Spain
            \\
           }
           \date{\today}

\titlerunning{Gaia white dwarfs in the Hercules stream}
\authorrunning{Torres et al.}

\offprints{S. Torres}

%_____________________________________________________________________

% \abstract{}{}{}{}{} 
% 5 {} token are mandatory
 
  \abstract
{}
{We analyzed the velocity space of the thin and thick-disk  {\it Gaia} white dwarf population within 100\,pc looking for signatures of the Hercules stellar stream. We aimed to identify those objects belonging to the Hercules stream and, by taking advantage of  white dwarf stars as reliable cosmochronometers,  to derive a first age distribution.}
{We applied a kernel density estimation to the $UV$ velocity space of white dwarfs.  For the region where a clear overdensity of stars was found, we created a 5-D space of dynamic variables. We applied a hierarchichal clustering method, {\tt HDBSCAN}, to this 5-D space, identifying those white dwarfs that share similar kinematic characteristics. Finally, under general assumptions and from their photometric properties,  we derived an age estimate for each object. }
{The Hercules stream was firstly revealed as an overdensity in the $UV$ velocity space of the thick-disk white dwarf population. Three substreams were then found: Hercules $a$ and Hercules $b$, formed by thick-disk stars with an age distribution peaked $4\,$Gyr in the past and extended to very old ages; and Hercules $c$, with a ratio of 65:35 thin:thick stars and a more uniform age distribution younger than 10\,Gyr}
{}


\maketitle
%
%________________________________________________________________

\section{Introduction}

Stars from the solar neighbourhood (the volume of the Galaxy up to a few hundred pc from the Sun) are far from presenting a uniform and homogeneous distribution of velocities. Additionally to the Galactic components of the thin- and thick-disk and the stellar halo, several kinematic structures left their imprint in the velocity space. The origin of these structures is under an intense debate and involves a large variety of hypotheses ranging from non-axisymmetric structural components (such as the bar and the spiral arm of the Galaxy) to cluster disruption or past accretion events \citep[for a thorough review see][]{Antoja2010}. In any case, they represent relevant signatures of the structure and dynamical evolution of the Galaxy.

Among them, one of the most prominent kinematic features is the Hercules stream, also known as the $U$-anomaly. The Hercules stream, first seen in \citet{Eggen1958} and later detected in multiple surveys such as Hipparcos, RAVE, LAMOST and {\it Gaia} \citep[e.g.][]{Dehnen1998, Antoja2012,Liang2017,GC2018},  is revealed as an elongated region in the $UV$-plane with a characteristic velocity moving away from the Galactic center, $U\approx-30-50$\kms and lagging behind the local standard of rest (LSR), $V\approx-60$\kms. Discarded the hypothesis of the disruption of a cluster as its origin, given their spread in ages and metallicities \citep[e.g.][]{Famaey2005,Antoja2008,Bovy2010}, the stream was more likely formed due to non-axisymmetries in the Galactic potential.
In particular, the Hercules stream has been linked to be caused by the
effects of the Outer Lindblad Resonance (OLR) of the Galactic bar for a long time \citep{Dehnen2000,Fux2001}. 
More recently, alternative origins for Hercules have been proposed given the independent evidence for a long slow-rotating bar in the Milky Way \citep{Wegg2015,Portail2015} that would place the OLR too far beyond the solar neighbourhood. Thus, Hercules stream  could be related to the corotation resonance of the bar \citep{PerezVillegas2017}, the 4:1 OLR of a slow bar \citep{Hunt2018}, and a combination of the effects of a slow bar and spiral arms \citep{Hattori2019}. This issue is far from being settled, and recent work still supporting the original explanation   \citep{Hunt2018a, Ramos2018,Fragkoudi2019}.

 While Hercules is the group with largest extension in the velocity space, multiple studies have revealed substructures within it. For instance, two overdensities at approximately the same rotation velocity but at different Galactocentric radial velocity are observed in \citet{Dehnen1998, Antoja2012}. Recently, {\it Gaia} data showed the splitting of the Hercules stream in two/three branches of approximately constant   Galactocentric radial velocity \citep{GC2018,Ramos2018}. 

On the other hand, white dwarfs are long living objects representing the most common evolutionary remnants of low- and intermediate-mass stars --i.e. those with  $M\,\lappr\,8\sim 11\,M_{\sun}$, \cite[e.g.][]{Siess2007}. Nuclear fusion reactions have ceased in white dwarf interiors, being the pressure due to the degenerate electrons the responsible to prevent the gravitational collapse of these compact objects. White dwarfs are, then, subjected to a long process of gravothermal cooling, and their characteristics are reasonably well understood from a theoretical point of view -- see, for instance, the review by \cite{Althaus2010a}. Hence, white dwarfs are promoted as ideal candidates to constrain the age and formation history of the different Galactic components \citep[e.g.][]{Fontaine2001,Reid2005,GB2016}.

However, to date, the use of the white dwarfs as Galactic tracers of the dynamic evolution has been limited. In particular, only a few modest studies of the white dwarf population have been dedicated to the search of stellar streams in the solar neighborhood \citep[e.g.][]{Fuchs2011} or to study the possible imprints of a merger episode in the Galactic disk \citep[e.g.][]{Torres2001}. Several reasons account for this: first, white dwarfs lack from radial velocity measurements, unless  optimal resolution spectra are available \cite[e.g.][]{Pauli2006,Anguiano17}.  This fact is a consequence of the broadening of the Balmer’s spectral lines, due to the large surface gravity characterizing white dwarf atmospheres. Second, also due to this huge gravitational pull, metals are sinked in the deep interiors of white dwarf envelopes, precluding from associating a metallicity value to these objects. Finally, the number of white dwarfs identified in volume-limited samples is rather small.

Fortunately, the advent of large data surveys such as {\it Gaia} has dramatically increased the number of known white dwarfs, thus providing statistically significant large complete samples. This fact, together with the potential of white dwarfs as cosmochronometers, opens the door to use these objects as reliable tracers of the dynamic Galactic evolution. 

In this Letter we aim to put into manifest the Hercules stream signature in the population of white dwarfs of the local neighborhood. Additionally, we aim to identify those white dwarfs belonging to the Hercules stream and to derive a first approach  to their age distribution.

%__________________________________________________________________

\section{The {\it Gaia} DR2 white dwarf 100\,pc sample}

The second data release of the {\it Gaia} mission has provided an unprecedented wealth of accurate astrometric and photometric information. In particular, nearly  $260,000$ objects have been identified as white dwarf candidates \citep{Gentile2019} up to a few kpc from the Sun. Moreover, \cite{Jimenez2018} claimed that the largest and nearly complete {\it Gaia} white dwarf sample extends up to 100\,pc from the Sun and contains close to $18,000$ objects. For this sample, and by means of  applying innovative techniques based on a Random Forest machine learning algorithm, \cite{Torres2019} have been able to classify the white dwarf population into its different  Galactic components: thin and thick-disk and stellar halo. The final classified  sample contains 12,227  thin-disk,  1,410  thick-disk  and  nearly 100 halo  white  dwarf  candidates. Heliocentric velocities are calculated in the standard Galactic coordinate velocity system, $(U,V,W)$, with $U$ positive towards the Galactic centre, once derived from the 5-parameter astrometric solution, $(\alpha,\delta,\varpi,\mu^*_{\alpha}, \mu_{\delta})$, provided by the {\it Gaia} measurements.
%______________________________________________________________

\section{Identifying white dwarfs in the Hercules stream}

\begin{figure}[t]
   %\resizebox{\hsize}{!}
%   {\includegraphics[width=0.97\columnwidth]{Thinthickdisk.eps}}
{\includegraphics[width=0.97\columnwidth]{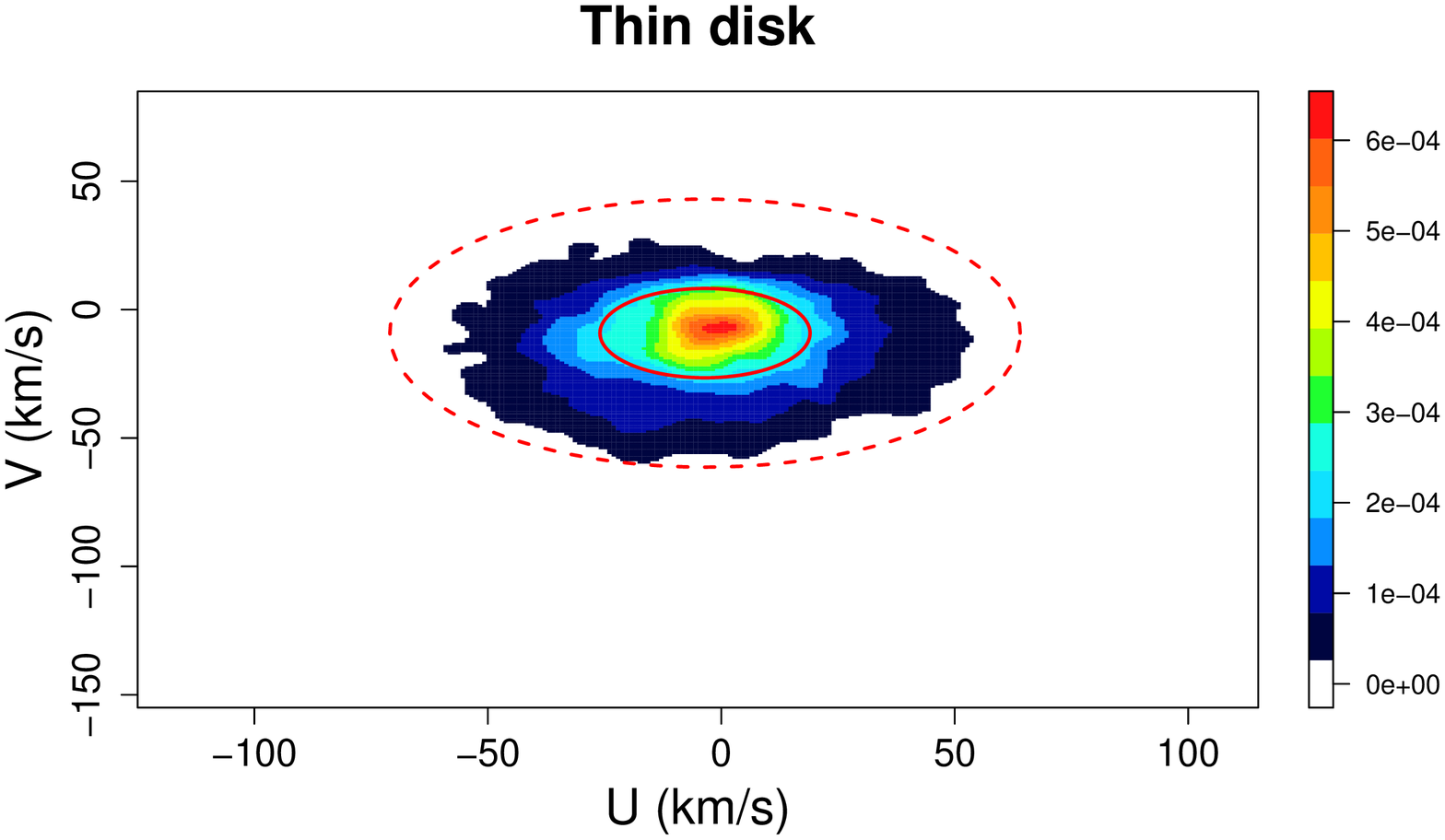}} 
{\includegraphics[width=0.97\columnwidth]{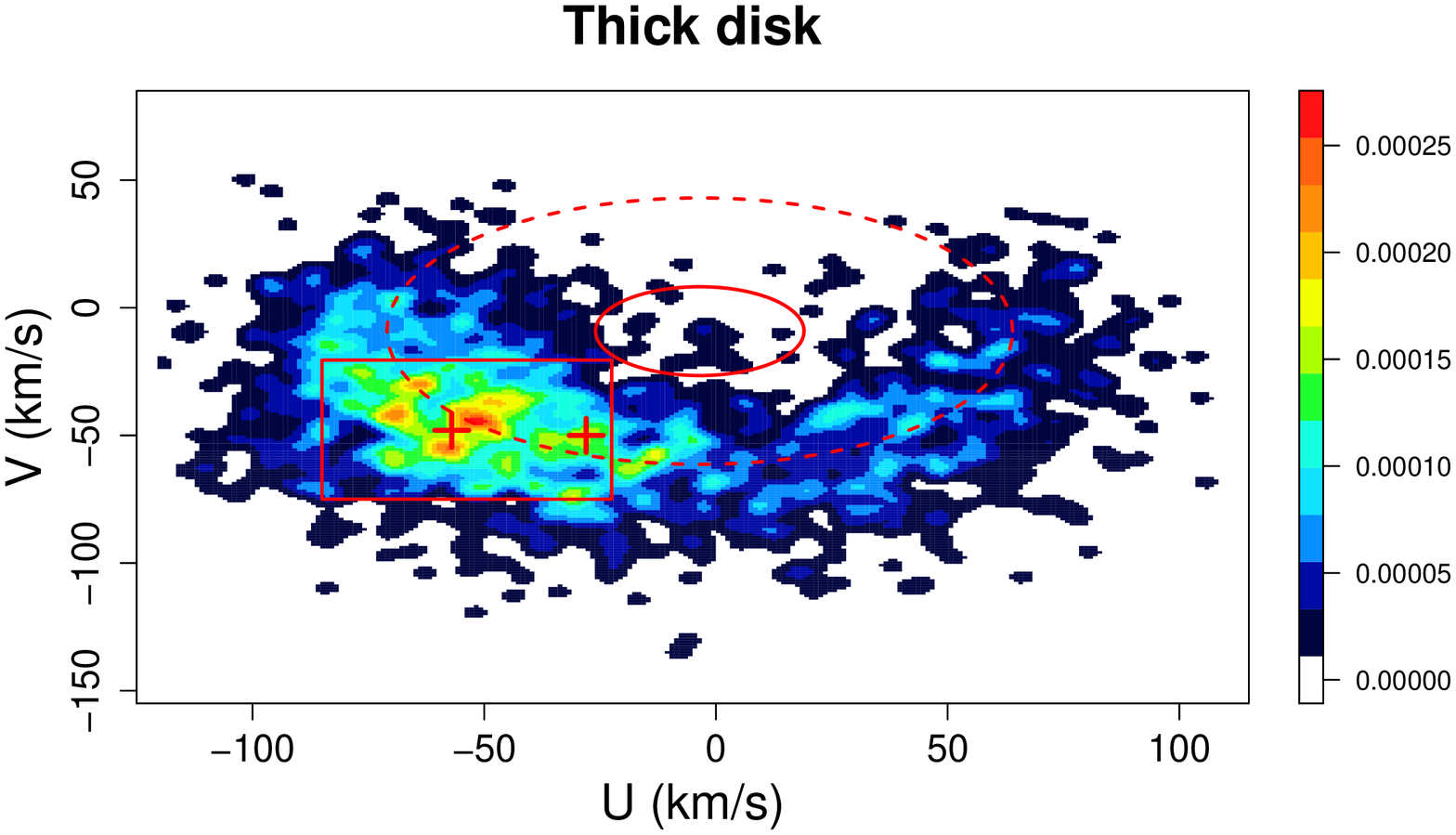}}
\caption{$UV$ density diagram for the thin (top panel) and thick-disk (bottom panel) white dwarf population within 100\,pc from the Sun. A clear overdensity is revealed in the thick-disk population (delimited by a red  rectangle). For comparative purposes, we show the thin-disk dispersion ellipsoid for $1\sigma$ (continuous line) and $3\sigma$ (dashed line) levels corresponding to the thin-disk population \citep{Torres2019} and we illustrate the number density (right scales) of white dwarf stars per $1\times1\,({\rm km/s})^2$. Also shown as red crosses are the location of the Hercules sub-streams from \citet{Antoja2012}. See text for details. }
\label{f:uvplane}
\end{figure}

The Hercules stream, as many other similar stellar streams, is firstly revealed as an overdensity of stars in the $UV$ plane \citep[e.g.][]{Dehnen2000}. Thus, as shown in Figure \ref{f:uvplane}, we start by analyzing the $UV$ space velocity plane by means of a density kernel estimation \citep{chen1997} for the 100\,pc {\it Gaia} thin and thick-disk white dwarf populations. No relevant feature departing from a Gaussian distribution appears when the thin-disk white dwarf population is depicted (top panel of Fig. \ref{f:uvplane}). In fact, the $U$ distribution (once corrected from the solar motion) is reasonably symmetric: $53\%$ to $47\%$ of objects with $U>0$ and $U<0$, respectively, in the region with $V<0$.  However, when  the white dwarf thick-disk population is  represented (bottom panel of Fig. \ref{f:uvplane}) a clear overdensity, breaking the $U$ symmetry, is revealed. Now, $68\%$ of these objects  present a negative $U$ velocity. This overdensity region (red rectangle), centered at around $(U,V)=(-55,-50)$\kms and with an extension of $(\Delta U,\Delta V)=(60,50)$\kms , presents a number of objects per (\kms)$^2$ roughly 3 times larger than the average density. The so called $U${\sl-anomaly} is thus clearly put into manifest in the $UV$ plane for the thick-disk distribution. 

It is worth mentioning that this feature is not a consequence of the lack of radial velocities in the white dwarf sample. Our Monte Carlo population synthesis analysis performed in \cite{Torres2019} reveals that the assumption of null radial velocity implies a reduction of the speed moduli of the stars. In particular, the $(U,V)$ components are expected to be reduced by $(3.7\pm17,7.2\pm18)$\kms and $(4.0\pm29,10.7\pm32)$\kms on average for the thin and thick-disk populations, respectively. However, this fact does not generate any asymmetry in their final component distributions. In the same sense, the incompleteness of the thick-disk white dwarf sample can not be argued as the cause of this peculiar overdensity. Thick-disk white dwarfs are extremely difficult to disentangle from thin-disk white dwarfs for low speed stars. This fact is the responsible of the characteristic {\sl croissant}-shape appearing in the thick-disk distribution, but again, it cannot justify the observed overdensity.

\begin{figure*}[t]
   %\resizebox{\hsize}{!}

      {\includegraphics[width=1.0\columnwidth, clip=true,trim=35 45 35 30]{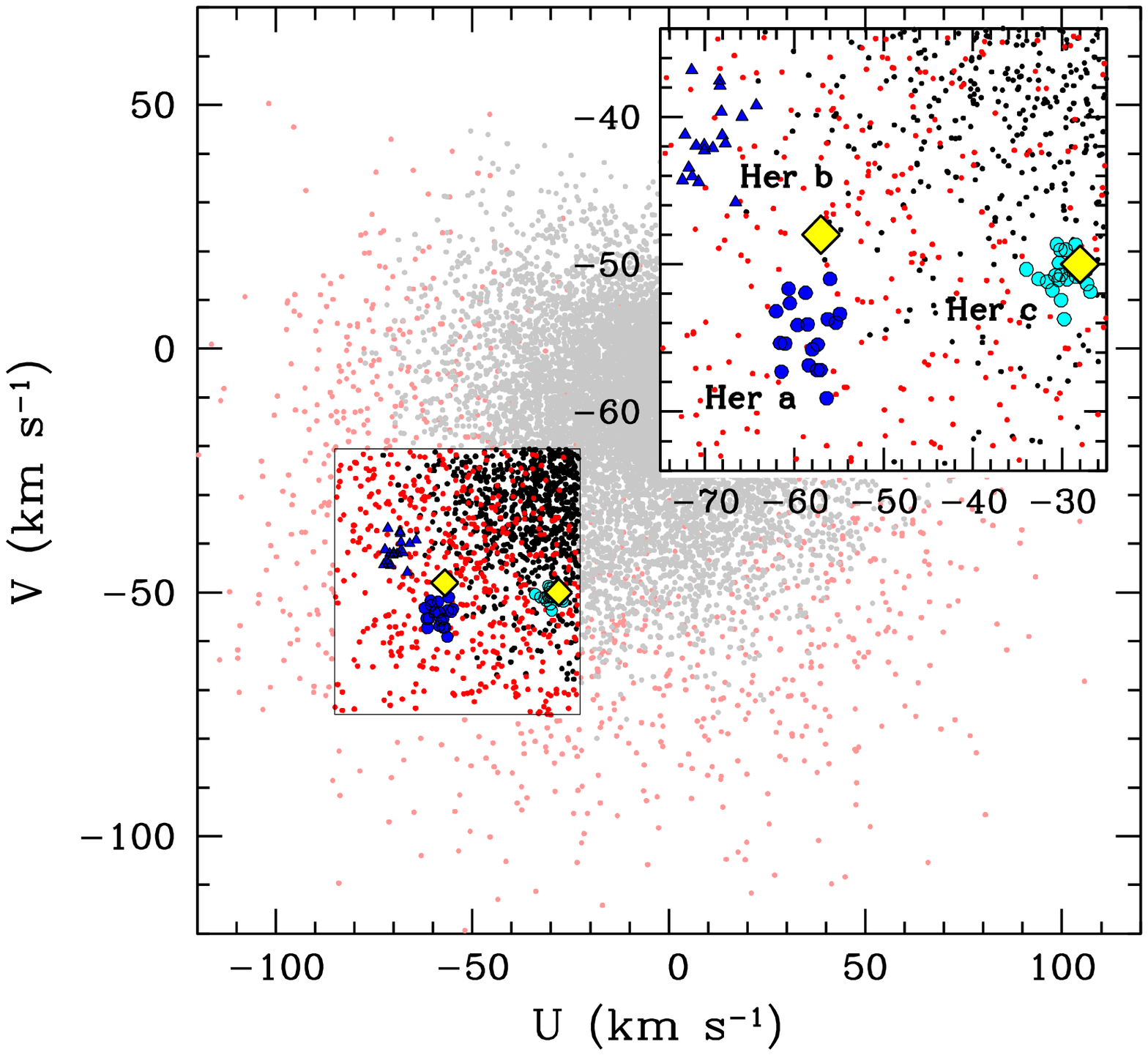}}
        {\includegraphics[width=1.0\columnwidth,clip=true,trim=35 45 35 20]{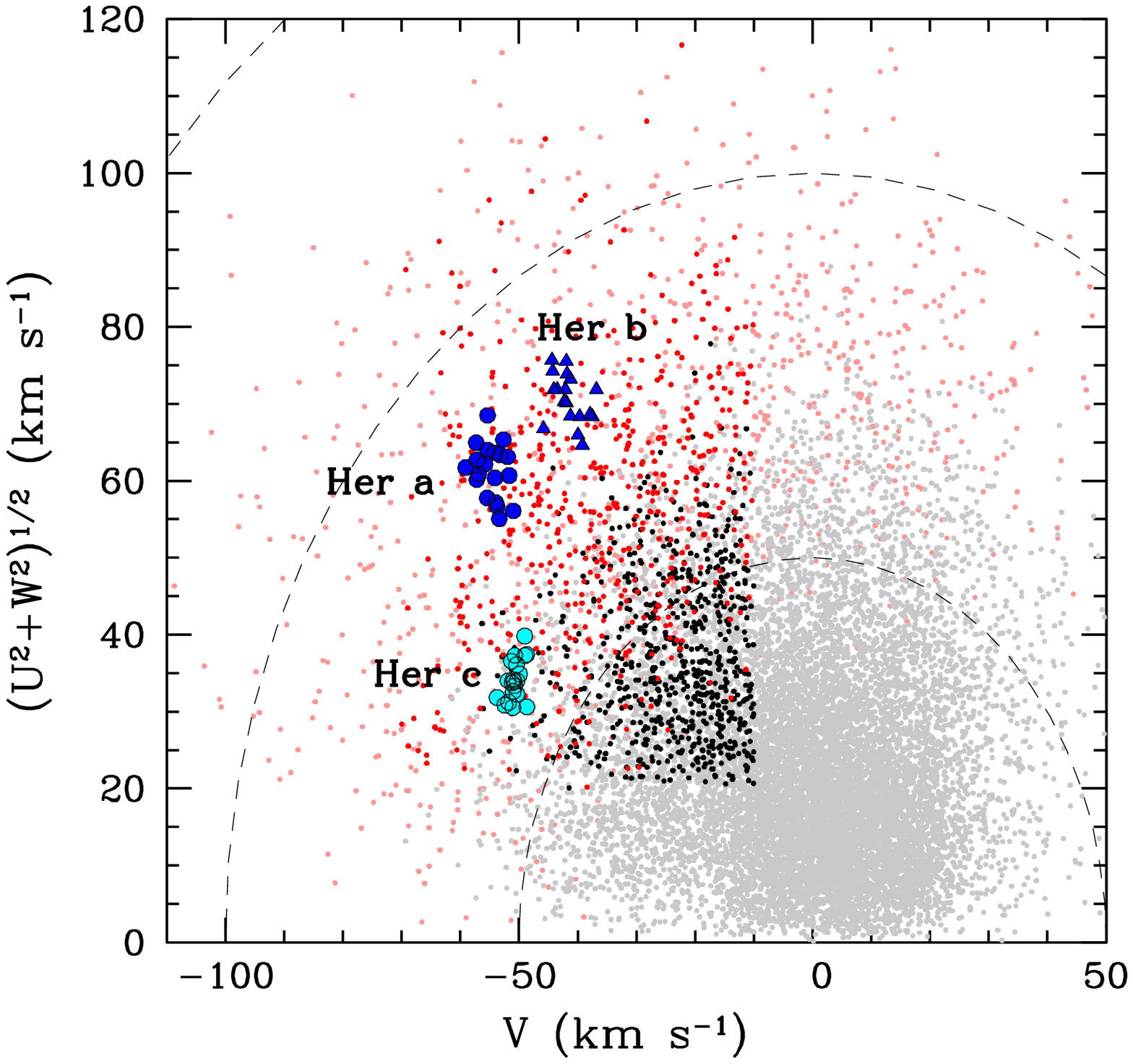}}  
\caption{$UV$ diagram (left panel) and Toomre diagram (right panel) for the thin  and thick-disk white dwarf population (black and red dots, respectively, and color emphasized those belonging to the selected overdensity region). Also plotted are the structures Her $a$ (blue circles), Her $b$ (blue triangles) and Her $c$ (cyan circles) found in this work}.  For comparative purposes,  we show  as yellow squares the locations in the $UV$ plane of the Hercules sub-streams from \cite{Antoja2012}.
\label{f:panels}
\end{figure*}

Once discarded any possible selection bias as its cause, the overdensity of white dwarfs found in the $UV$ plane is revealed as a kinematic feature different in nature to the thin or thick-disk populations. Moreover, the location within the $UV$ plane is in perfect agreement with previous identifications of the Hercules stream -- see the bottom panel of Fig. \ref{f:uvplane}. We can conclude then that the Hercules stream signature is present in the kinematics of the white dwarf population.

We now aim to identify those white dwarfs that are genuine members of the Hercules stream.  The loci of this stream within the $UV$ plane is superimposed to the standard population of thick-disk white dwarfs and also to  high speed thin-disk stars. For this reason, we need to extend our analysis in order to disentangle the different populations. Given the lack of radial velocity and metallicity measurements, we are compelled to extract the maximum possible information contained in the tangential velocity of each white dwarf. To this end, the following procedure has been carried out. First, we create a 5D-space with those variables that are relevant for disentangling stellar kinematic populations. In our case, the chosen variables are: the $U$ and $V$ Galactic velocity components; the modulus or peculiar velocity of the star, $V_{\rm pec}\equiv(U^2+V^2+W^2)^{1/2}$; the velocity perpendicular to the Galactic rotation as defined in the Toomre diagram, $V_{\rm Toomre}\equiv(U^2+W^2)^{1/2}$; and $V_{\Delta E}=(U^2+2V^2)^{1/2}$ which is derived from the integral of motion and it is related to the eccentricity, $e$, of the orbit through $V_{\Delta E}=\sqrt{2}ev_{\rm c}$, being $v_{\rm c}$ the Galactic velocity assuming a flat rotation model. This set of variables has been widely used in the search of stellar kinematic structures \cite[e.g.][]{Bensby2007,Fuchs2011} and provides a complete set of dynamical properties.

Our second step is to apply a machine learning technique in order to identify a genuine group of Hercules stream white dwarfs, since it is expected that members of a kinematic structure  share similar characteristics in our 5-D space. Among the machine learning techniques, several examples of unsupervised density-based clustering multipurpose methods widely used in astrophysics are available. Among  them, the most popular are {\tt DBSCAN}, {\tt HDBSCAN} and {\tt OPTICS}, \citep[see, for instance,] [for a thorough description of the algorithms and references there in]{Canovas2019}. Here we choose {\tt HDBSCAN}, which represents an extension of {\tt DBSCAN} and improves its performance by implementing a hierarchical clustering strategy. The number of  hyperparameters (free parameters introduced by the user) needed by the {\tt HDBSCAN} algorithm is minimal. We just adopt a value of the hyperparameter {\sl mPts} (minimum number of objects to form a cluster) in the range 10-30 and a probability to belong to a cluster larger than $90\%$. The first criterion ensures a minimum physical significance of the cluster, while the second one guarantees that all members of a particular cluster share similar properties. 

Once our 5-D space is normalized -- following a  standard scaler and using an euclidean metric --, we apply the {\tt HDBSCAN} algorithm to the $UV$ overdensity region (red rectangle in the bottom panel of Fig. \ref{f:uvplane}) to our thin plus thick-disk white dwarf population. The algorithm found 4 clusters or groups. The main group, formed by close to 400 objects, contains mainly field thin-disk white dwarfs that lie within the selected region. The other 3 groups, the ones we are interested in, are formed by close to 20 stars each one, revealing kinematic characteristics different from that of the thick and thin-disk population. These 3 groups, along with the white dwarfs of the overdensity selected region for different combinations of our 5-D space variable, are represented in Figure \ref{f:panels}. The first of these 3 groups  obtained by {\tt HDBSCAN} (blue circles) is located  in the $UV$ plane (left panel of Fig.\ref{f:panels}) at $(U,V)=(-58.5, -54.7)$\kms and contains 19 white dwarfs. The second group (blue triangles), close to the first one, is located at $(U,V)=(-69.3, -41.4)$\kms and contains 18 objects. Finally, the third group (cyan circles) is formed by 20 stars centered at $(U,V)=(-29.9, -50.7)$\kms.  We will call these groups Hercules (Her) $a$, $b$ and $c$, respectively. The first relevant conclusion here is that these groups are in excellent agreement with the Hercules stream location found in the literature. Although the Hercules stream was initially discovered as a diffuse elongated region in the $UV$ space, later analyses claim that the Hercules stream is formed by two substreams called Hercules I and II  \citep{Antoja2012,Bobylev2016}.  Recent studies based on several million FGK stars provided by the {\it Gaia}-DR2 analysis \citep{Gaia2018} reveal, instead, Hercules substreams as thin-arch structures \citep{Ramos2018,Li2019}. Our groups Her $a$ and $b$ are in agreement with the Hercules I substructure, while our group  Her $c$  perfectly matches Hercules II. On the other hand,  our Her $b$ and $c$ are consistent with structure A9 and Her $a$ with A8 found in \cite{Ramos2018}. A deeper study (including, for instance, radial velocities) is needed to ascertain the ultimately origin of our three groups. Meanwhile, we will treat them separately. In Table \ref{t:loc} we summarize the results.\footnote{A complete list of the white dwarfs belonging to each of the three groups identified in this work is available upon request.}

   \begin{table}
      \caption[]{Hercules sub-stream locations.}
         \label{t:loc}
         \begin{center}
     {\small
\begin{tabular}{ccc}  
            \hline
            \noalign{\smallskip}
            Source      &  Sub-  & $(U,V)$  \\
                        &   stream & (\kms) \\
            \noalign{\smallskip}
            \hline
            \noalign{\smallskip}
            This work  & Her $a$ & $(-59\pm2,-55\pm3)$     \\
            This work  & Her $b$ & $(-69\pm2,-41\pm2)$  \\
            This work  & Her $c$ & $(-30\pm2,-51\pm1)$    \\
            \cite{Antoja2012}  & Her I &$(-57,-48)$     \\
            \cite{Antoja2012}  & Her II & $(-28,-50)$     \\
             \cite{Bobylev2016}  & Her I & $(-57,-48)$    \\
            \cite{Bobylev2016}  & Her II & $(-35,-55)$    \\              \cite{Ramos2018}  & A8 &  $\sim([-30,-70],-35)$    \\  
            \cite{Ramos2018}  &  A9 &  $\sim ([-50,+10],-50)$    \\

            \noalign{\smallskip}
            \hline
\end{tabular}
}
\end{center}
   \end{table}

It is worth mentioning  that our Her $a$ structure is exclusively formed by thick-disk stars, while for Her $b$, all but one (5\%)  belong to the thick-disk population. Conversely, our Her $c$ substructure is formed by 13(65\%) thin-disk and 7(35\%) thick-disk  white dwarfs. These results are also in agreement with the spread in metallicities found in the Hercules stream \citep{Hattori2019}.

Although one could argue that  the effects of the bar are mostly constrained to the thin-disk population under the claim that orbits reaching high altitudes in the plane would not be that much perturbed, simulations have shown that the bar induces a Hercules-like structure on the thick disk comparable to that of the thin-disk. Thus, the existence of Hercules in the thick disk population is, from the dynamical point of view, totally plausible in the scenario of the bar resonant effects. The first observational evidence that the effects of the bar are present also in the thick-disk was shown in \citet{Antoja2012} and, more recently,  \citet{Koppelman2018} have also identified an asymmetry in the $U$ velocity tail of the thick-disk that could have the same origin.

\begin{figure*}[t]
   %\resizebox{\hsize}{!}

      {\includegraphics[width=0.5\columnwidth, clip=true,trim=62 45 62 30]{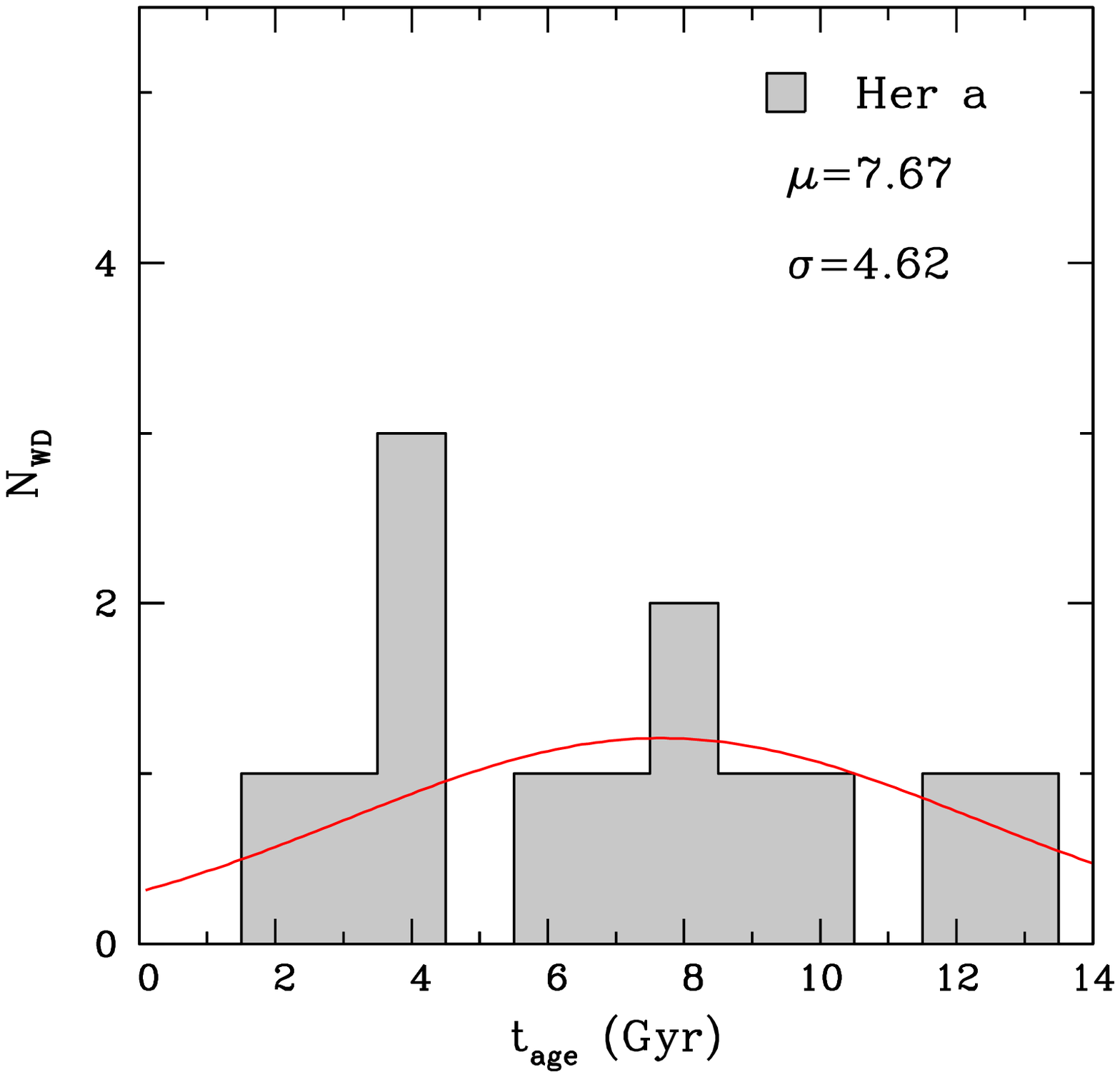}}
      {\includegraphics[width=0.5\columnwidth, clip=true,trim=62 45 62 30]{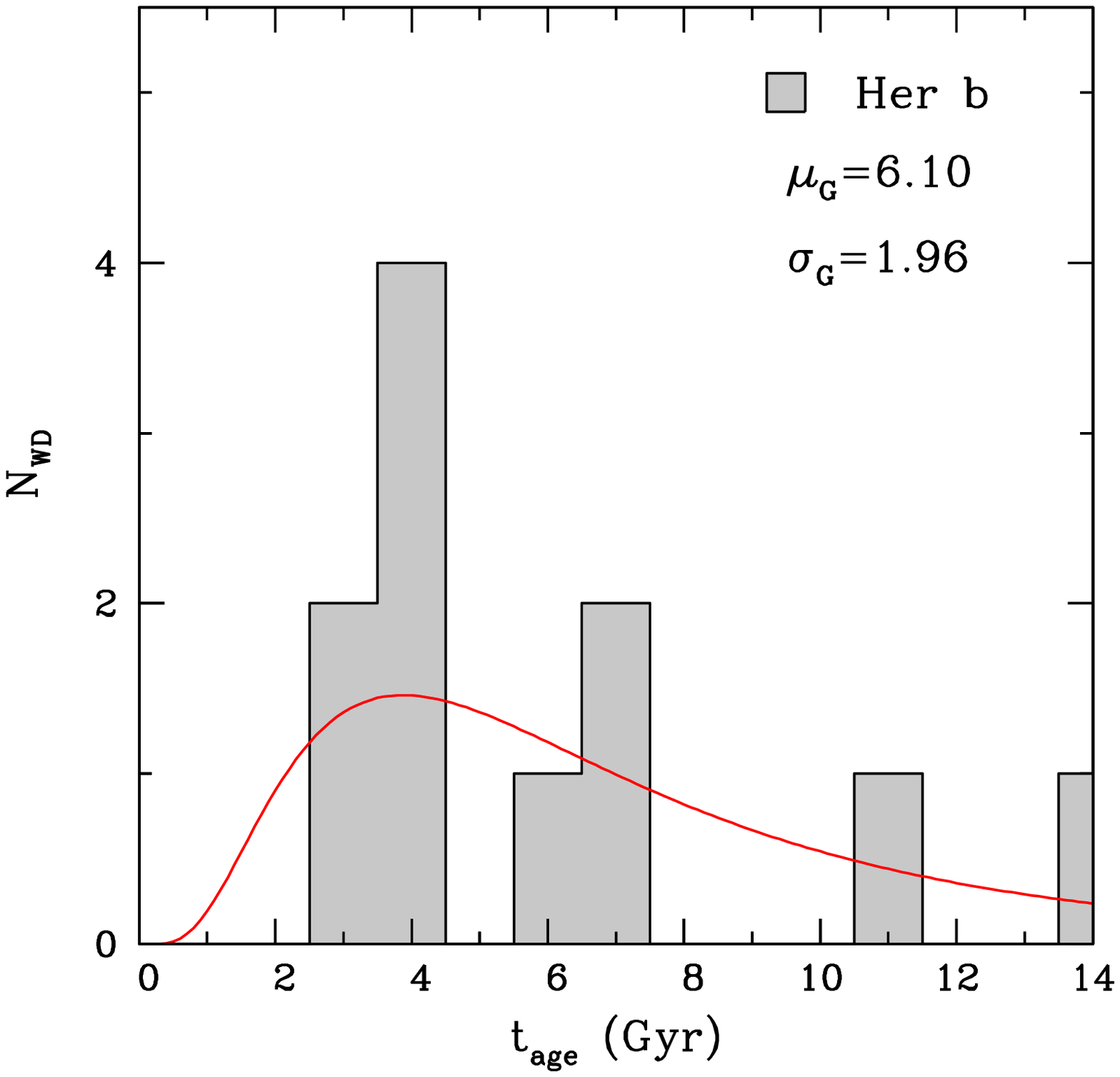}}
      {\includegraphics[width=0.5\columnwidth, clip=true,trim=62 45 62 30]{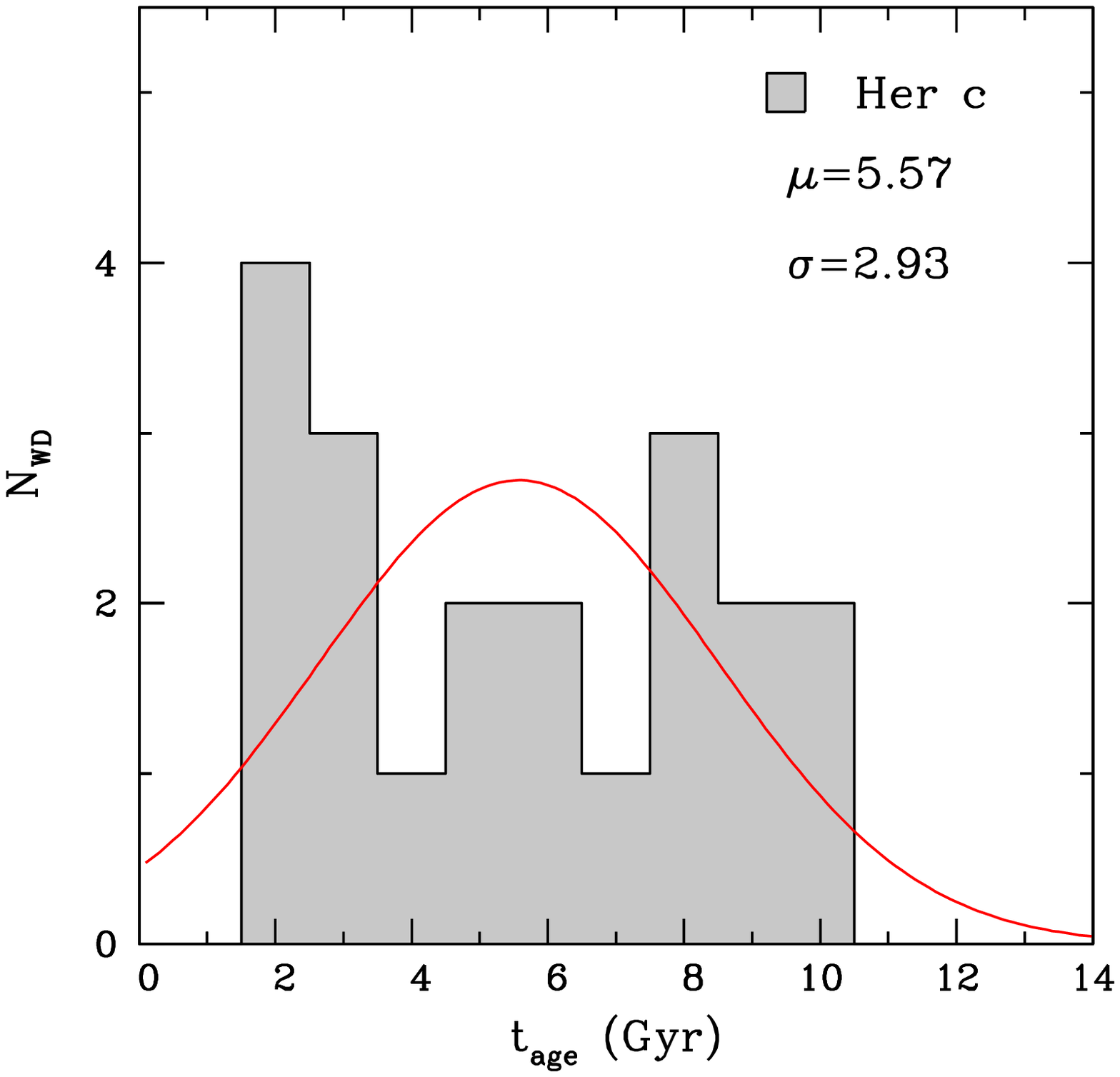}}
      {\includegraphics[width=0.5\columnwidth, clip=true,trim=62 45 62 30]{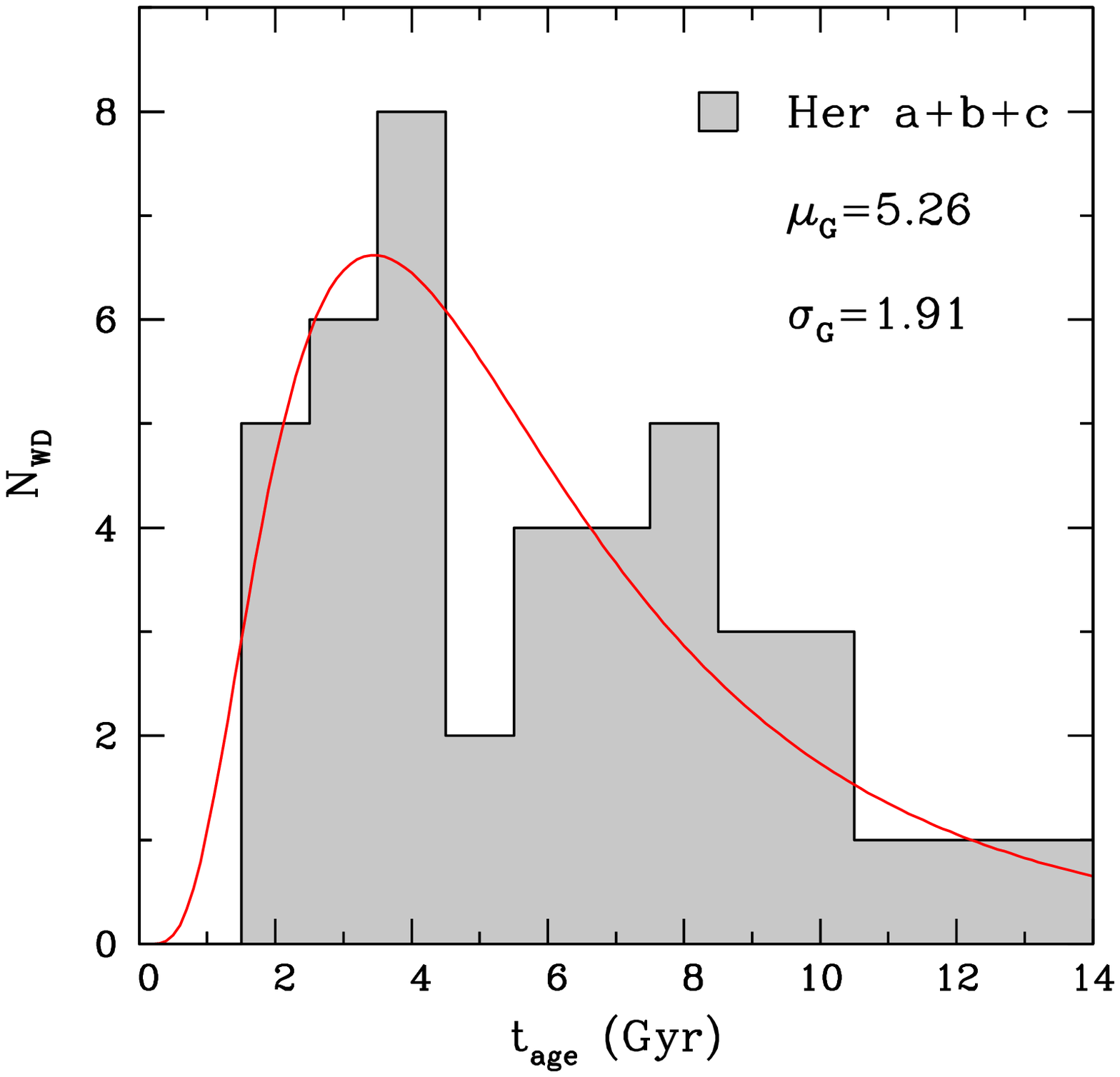}}

\caption{ Age distribution for each of the three Hercules sub-stream identified in this work and for all of them considered together (right panel). The sub-stream Her $b$  presents a maximum at around $4\,$Gyr with an extended long tail up to very old ages. Her $a$ is similar, but with a less clear peak. Sub-stream Her $c$ appears as a more uniform distribution between 2 and 10 Gyr. Also plotted as red lines are analytical normal fits for Her $a$ and $c$  and a log-normal fit for Her $c$ and the entire distribution. The corresponding mean $\mu$ and  deviation $\sigma$, the geometric mean $\mu_G$ and the multiplicative deviation $\sigma_G$, are depicted in each panel.}
\label{f:ages}
\end{figure*}

%______________________________________________________________

\section{Age distribution estimation}

A first estimation of the ages of the white dwarfs so far identified in the Hercules stream is performed. Due to the lack of spectroscopic information we are compelled to adopt some assumptions. First, we assume hydrogen-rich atmospheres (also known as DA white dwarfs) for all the stars in our sample. Second, we adopt a solar metallicity value, Z=0.01, for thin-disk white dwarfs. Although a small dispersion around the solar value is expected \citep[e.g.][]{Tononi2019}, discrepancies in the age determinations are negligible once the  final age distribution is binned in intervals of $1\,$Gyr. Similarly, a subsolar value, Z=0.001 is adopted for the thick-disk white dwarfs. From the parallax and photometry measurements provided by {\it Gaia} we derive the absolute magnitude, $M_{\rm G}$, and the color $G_{\rm BP}-G_{\rm RP}$ for each star of our sample. Once a set of cooling tracks is adopted (we used those of \citealt{Althaus2015,Camisassa2016,2018arXiv180703894C},  which encompass a full range of masses and progenitor metallicities and provide a full set of self-consistent cooling sequences) we obtain a robust estimate of the total age (white dwarf cooling time plus progenitor life-time) for each object of our Hercules sample. Objects with masses below $0.53\,{\rm M_{\odot}}$ are discarded  since they probably belong to binary systems in which a common envelope episode has occurred and, therefore, their ages are not possible to trace back \citep[e.g.][]{Rebassa2011}.

In Figure \ref{f:ages} we show the age distribution for our white dwarfs identified in the Her $a$, $b$ and $c$  streams and also  the distribution belonging to the three streams together (right panel).  Additionally, a Kolmogorov-Smirnov (K--S) test between the age distributions reveals that the null hypothesis (i.e. that two distributions have the same origin) can not be rejected with a probability, $P$, of $\sim0.70$, when Her $a$ and $c$ or Her $b$ and $c $ are compared. However, when Her $a$ and $b$ are compared the probability to have the same origin decreases to $P=0.44$. Our  K-S test also shows (see Fig.  \ref{f:ages}) that Her $a$ and Her $c$ are consistent with a normal distribution ($P=0.94$ and $P=0.86$), while the age distribution of Her $c$ is unlikely normally distributed ($P=0.17$), but consistent with a log-normal distribution ($P=0.81$). When the entire population of Hercules  white dwarfs is considered, a log-normal distribution peaked 4\,Gyr in the past with a secondary slight peak at 8\,Gyr extended to old ages seems a reasonable fit ($P=0.61$), disregarding a double-gaussian distribution ($P=0.13$).   Although these distributions should be understood as preliminary guesses, the extended ages found here are in agreement with the dynamical origin of the Hercules stream, disregarding any cluster disruption hypothesis.  

%______________________________________________________________

\section{Conclusions}

We have revealed the imprint of the Hercules stream in the space velocity of white dwarfs in the solar neighborhood. In particular, we analyzed the recent {\it Gaia} white dwarf population, which presents a valuable wealth of accurate photometric and astrometric data and a nearly complete sample within 100 pc. The analysis of the  white dwarf $UV$ plane puts into manifest an overdensity of objects in agreement with previous observed signatures of the Hercules stream  in main-sequence stars by other surveys like Hipparcos, RAVE or  LAMOST.  

Taking advantage of an advanced hierarchical clustering algorithm, {\tt HDBSCAN}, applied to a 5-D space of kinematic variables, we were able to identify those white dwarfs that are genuine members of the Hercules stream.  Three main substreams, Her $a$, $b$ and $c$, located at  $(U,V)=(-59, -55)$\kms, $(-69, -41)$\kms and $(-30, -51)$\kms respectively, were identified. The first two are practically formed by thick-disk white dwarfs. The third one is a mixture of $65:35\%$ of thin and thick-disk white dwarfs, respectively. Moreover, a first guess of their age distribution shows that Her $b$ presents a maximum 4\,Gyr ago and extends up to very old ages, similar to Her $a$. However  Her $c$ depicts a more uniform distribution of objects between 2 and 10 Gyr. Although the nature and origin of the Hercules stream still remain as opens questions, we believe that  the Hercules white dwarf sample identified in this work can provide very valuable information for  its clarification.
 
\begin{acknowledgements}
This work  was partially supported by the MINECO grant AYA\-2017-86274-P and the Ram\'on y Cajal programme RYC-2016-20254, by  the AGAUR,  and by grant G149 from University of La Plata.
\end{acknowledgements}

%-------------------------------------------------------------------

\bibliographystyle{aa}
\bibliography{wdhs}

\begin{thebibliography}{43}
\expandafter\ifx\csname natexlab\endcsname\relax\def\natexlab#1{#1}\fi

\bibitem[{{Althaus} {et~al.}(2015){Althaus}, {Camisassa}, {Miller Bertolami},
  {C{\'o}rsico}, \& {Garc{\'{\i}}a-Berro}}]{Althaus2015}
{Althaus}, L.~G., {Camisassa}, M.~E., {Miller Bertolami}, M.~M., {C{\'o}rsico},
  A.~H., \& {Garc{\'{\i}}a-Berro}, E. 2015, \aap, 576, A9

\bibitem[{{Althaus} {et~al.}(2010){Althaus}, {C{\'o}rsico}, {Isern}, \&
  {Garc{\'{\i}}a-Berro}}]{Althaus2010a}
{Althaus}, L.~G., {C{\'o}rsico}, A.~H., {Isern}, J., \& {Garc{\'{\i}}a-Berro},
  E. 2010, \aapr, 18, 471

\bibitem[{{Anguiano} {et~al.}(2017){Anguiano}, {Rebassa-Mansergas},
  {Garc{\'{\i}}a-Berro}, {Torres}, {Freeman}, \& {Zwitter}}]{Anguiano17}
{Anguiano}, B., {Rebassa-Mansergas}, A., {Garc{\'{\i}}a-Berro}, E., {et~al.}
  2017, \mnras, 469, 2102

\bibitem[{{Antoja} {et~al.}(2008){Antoja}, {Figueras}, {Fern{\'a}ndez}, \&
  {Torra}}]{Antoja2008}
{Antoja}, T., {Figueras}, F., {Fern{\'a}ndez}, D., \& {Torra}, J. 2008, \aap,
  490, 135

\bibitem[{{Antoja} {et~al.}(2010){Antoja}, {Figueras}, {Torra}, {Valenzuela},
  \& {Pichardo}}]{Antoja2010}
{Antoja}, T., {Figueras}, F., {Torra}, J., {Valenzuela}, O., \& {Pichardo}, B.
  2010, Lecture Notes and Essays in Astrophysics, 4, 13

\bibitem[{{Antoja} {et~al.}(2012){Antoja}, {Helmi}, {Bienayme},
  {Bland-Hawthorn}, {Famaey}, {Freeman}, {Gibson}, {Gilmore}, {Grebel},
  {Minchev}, {Munari}, {Navarro}, {Parker}, {Reid}, {Seabroke}, {Siebert},
  {Siviero}, {Steinmetz}, {Williams}, {Wyse}, \& {Zwitter}}]{Antoja2012}
{Antoja}, T., {Helmi}, A., {Bienayme}, O., {et~al.} 2012, \mnras, 426, L1

\bibitem[{{Bensby} {et~al.}(2007){Bensby}, {Oey}, {Feltzing}, \&
  {Gustafsson}}]{Bensby2007}
{Bensby}, T., {Oey}, M.~S., {Feltzing}, S., \& {Gustafsson}, B. 2007, \apjl,
  655, L89

\bibitem[{{Bobylev} \& {Bajkova}(2016)}]{Bobylev2016}
{Bobylev}, V.~V. \& {Bajkova}, A.~T. 2016, Astronomy Letters, 42, 90

\bibitem[{{Bovy} \& {Hogg}(2010)}]{Bovy2010}
{Bovy}, J. \& {Hogg}, D.~W. 2010, \apj, 717, 617

\bibitem[{{Camisassa} {et~al.}(2018){Camisassa}, {Althaus}, {C{\'o}rsico}, {De
  Ger{\'o}nimo}, {Miller Bertolami}, {Novarino}, {Rohrmann}, {Wachlin}, \&
  {Garc{\'{\i}}a--Berro}}]{2018arXiv180703894C}
{Camisassa}, M.~E., {Althaus}, L.~G., {C{\'o}rsico}, A.~H., {et~al.} 2018,
  arXiv e-prints [\eprint[arXiv]{1807.03894}]

\bibitem[{{Camisassa} {et~al.}(2016){Camisassa}, {Althaus}, {C{\'o}rsico},
  {Vinyoles}, {Serenelli}, {Isern}, {Miller Bertolami}, \&
  {Garc{\'{\i}}a-Berro}}]{Camisassa2016}
{Camisassa}, M.~E., {Althaus}, L.~G., {C{\'o}rsico}, A.~H., {et~al.} 2016,
  \apj, 823, 158

\bibitem[{{C{\'a}novas} {et~al.}(2019){C{\'a}novas}, {Cantero}, {Cieza},
  {Bombrun}, {Lammers}, {Mer{\'{\i}}n}, {Mora}, {Ribas}, \&
  {Ru{\'{\i}}z-Rodr{\'{\i}}guez}}]{Canovas2019}
{C{\'a}novas}, H., {Cantero}, C., {Cieza}, L., {et~al.} 2019, arXiv e-prints
  [\eprint[arXiv]{1902.07600}]

\bibitem[{{Chen} {et~al.}(1997){Chen}, {Asiain}, {Figueras}, \&
  {Torra}}]{chen1997}
{Chen}, B., {Asiain}, R., {Figueras}, F., \& {Torra}, J. 1997, \aap, 318, 29

\bibitem[{{Dehnen}(1998)}]{Dehnen1998}
{Dehnen}, W. 1998, \aj, 115, 2384

\bibitem[{{Dehnen}(2000)}]{Dehnen2000}
{Dehnen}, W. 2000, \aj, 119, 800

\bibitem[{{Eggen}(1958)}]{Eggen1958}
{Eggen}, O.~J. 1958, \mnras, 118, 154

\bibitem[{{Famaey} {et~al.}(2005){Famaey}, {Jorissen}, {Luri}, {Mayor}, {Udry},
  {Dejonghe}, \& {Turon}}]{Famaey2005}
{Famaey}, B., {Jorissen}, A., {Luri}, X., {et~al.} 2005, \aap, 430, 165

\bibitem[{{Fontaine} {et~al.}(2001){Fontaine}, {Brassard}, \&
  {Bergeron}}]{Fontaine2001}
{Fontaine}, G., {Brassard}, P., \& {Bergeron}, P. 2001, \pasp, 113, 409

\bibitem[{{Fragkoudi} {et~al.}(2019){Fragkoudi}, {Katz}, {Trick}, {White}, {Di
  Matteo}, {Sormani}, {Khoperskov}, {Haywood}, {Hall{\'e}}, \&
  {G{\'o}mez}}]{Fragkoudi2019}
{Fragkoudi}, F., {Katz}, D., {Trick}, W., {et~al.} 2019, \mnras, 1815

\bibitem[{{Fuchs} \& {Dettbarn}(2011)}]{Fuchs2011}
{Fuchs}, B. \& {Dettbarn}, C. 2011, \aj, 141, 5

\bibitem[{{Fux}(2001)}]{Fux2001}
{Fux}, R. 2001, \aap, 373, 511

\bibitem[{{Gaia Collaboration} {et~al.}(2018{\natexlab{a}}){Gaia
  Collaboration}, {Katz}, {Antoja}, {Romero-G{\'o}mez}, {Drimmel}, {Reyl{\'e}},
  {Seabroke}, {Soubiran}, {Babusiaux}, \& {Di Matteo}}]{GC2018}
{Gaia Collaboration}, {Katz}, D., {Antoja}, T., {et~al.} 2018{\natexlab{a}},
  \aap, 616, A11

\bibitem[{{Gaia Collaboration} {et~al.}(2018{\natexlab{b}}){Gaia
  Collaboration}, {Katz}, {Antoja}, {Romero-G{\'o}mez}, {Drimmel}, {Reyl{\'e}},
  {Seabroke}, {Soubiran}, {Babusiaux}, {Di Matteo}, \& et~al.}]{Gaia2018}
{Gaia Collaboration}, {Katz}, D., {Antoja}, T., {et~al.} 2018{\natexlab{b}},
  \aap, 616, A11

\bibitem[{{Garc{\'{\i}}a-Berro} \& {Oswalt}(2016)}]{GB2016}
{Garc{\'{\i}}a-Berro}, E. \& {Oswalt}, T.~D. 2016, \nar, 72, 1

\bibitem[{{Gentile Fusillo} {et~al.}(2019){Gentile Fusillo}, {Tremblay},
  {G{\"a}nsicke}, {Manser}, {Cunningham}, {Cukanovaite}, {Hollands}, {Marsh},
  {Raddi}, {Jordan}, {Toonen}, {Geier}, {Barstow}, \& {Cummings}}]{Gentile2019}
{Gentile Fusillo}, N.~P., {Tremblay}, P.-E., {G{\"a}nsicke}, B.~T., {et~al.}
  2019, \mnras, 482, 4570

\bibitem[{{Hattori} {et~al.}(2019){Hattori}, {Gouda}, {Tagawa}, {Sakai},
  {Yano}, {Baba}, \& {Kumamoto}}]{Hattori2019}
{Hattori}, K., {Gouda}, N., {Tagawa}, H., {et~al.} 2019, \mnras, 484, 4540

\bibitem[{{Hunt} \& {Bovy}(2018)}]{Hunt2018}
{Hunt}, J. A.~S. \& {Bovy}, J. 2018, \mnras, 477, 3945

\bibitem[{{Hunt} {et~al.}(2018){Hunt}, {Bovy}, {P{\'e}rez-Villegas},
  {Holtzman}, {Sobeck}, {Chojnowski}, {Santana}, {Palicio}, {Wegg}, {Gerhard},
  {Almeida}, {Bizyaev}, {Fernandez-Trincado}, {Lane}, {Longa-Pe{\~n}a},
  {Majewski}, {Pan}, \& {Roman-Lopes}}]{Hunt2018a}
{Hunt}, J. A.~S., {Bovy}, J., {P{\'e}rez-Villegas}, A., {et~al.} 2018, \mnras,
  474, 95

\bibitem[{{Jim{\'e}nez-Esteban} {et~al.}(2018){Jim{\'e}nez-Esteban}, {Torres},
  {Rebassa-Mansergas}, {Skorobogatov}, {Solano}, {Cantero}, \&
  {Rodrigo}}]{Jimenez2018}
{Jim{\'e}nez-Esteban}, F.~M., {Torres}, S., {Rebassa-Mansergas}, A., {et~al.}
  2018, \mnras, 480, 4505

\bibitem[{{Koppelman} {et~al.}(2018){Koppelman}, {Helmi}, \&
  {Veljanoski}}]{Koppelman2018}
{Koppelman}, H., {Helmi}, A., \& {Veljanoski}, J. 2018, \apjl, 860, L11

\bibitem[{{Li} \& {Shen}(2019)}]{Li2019}
{Li}, Z.-Y. \& {Shen}, J. 2019, arXiv e-prints, arXiv:1904.03314

\bibitem[{{Liang} {et~al.}(2017){Liang}, {Zhao}, {Oswalt}, {Chen}, {Zhang}, \&
  {Zhao}}]{Liang2017}
{Liang}, X.~L., {Zhao}, J.~K., {Oswalt}, T.~D., {et~al.} 2017, \apj, 844, 152

\bibitem[{{Pauli} {et~al.}(2006){Pauli}, {Napiwotzki}, {Heber}, {Altmann}, \&
  {Odenkirchen}}]{Pauli2006}
{Pauli}, E.-M., {Napiwotzki}, R., {Heber}, U., {Altmann}, M., \& {Odenkirchen},
  M. 2006, \aap, 447, 173

\bibitem[{{P{\'e}rez-Villegas} {et~al.}(2017){P{\'e}rez-Villegas}, {Portail},
  {Wegg}, \& {Gerhard}}]{PerezVillegas2017}
{P{\'e}rez-Villegas}, A., {Portail}, M., {Wegg}, C., \& {Gerhard}, O. 2017,
  \apj, 840, L2

\bibitem[{{Portail} {et~al.}(2015){Portail}, {Wegg}, {Gerhard}, \&
  {Martinez-Valpuesta}}]{Portail2015}
{Portail}, M., {Wegg}, C., {Gerhard}, O., \& {Martinez-Valpuesta}, I. 2015,
  \mnras, 448, 713

\bibitem[{{Ramos} {et~al.}(2018){Ramos}, {Antoja}, \& {Figueras}}]{Ramos2018}
{Ramos}, P., {Antoja}, T., \& {Figueras}, F. 2018, \aap, 619, A72

\bibitem[{{Rebassa-Mansergas} {et~al.}(2011){Rebassa-Mansergas}, {Nebot
  G{\'o}mez-Mor{\'a}n}, {Schreiber}, {Girven}, \& {G{\"a}nsicke}}]{Rebassa2011}
{Rebassa-Mansergas}, A., {Nebot G{\'o}mez-Mor{\'a}n}, A., {Schreiber}, M.~R.,
  {Girven}, J., \& {G{\"a}nsicke}, B.~T. 2011, \mnras, 413, 1121

\bibitem[{{Reid}(2005)}]{Reid2005}
{Reid}, I.~N. 2005, \araa, 43, 247

\bibitem[{{Siess}(2007)}]{Siess2007}
{Siess}, L. 2007, \aap, 476, 893

\bibitem[{{Tononi} {et~al.}(2019){Tononi}, {Torres}, {Garc{\'\i}a-Berro},
  {Camisassa}, {Althaus}, \& {Rebassa-Mansergas}}]{Tononi2019}
{Tononi}, J., {Torres}, S., {Garc{\'\i}a-Berro}, E., {et~al.} 2019, arXiv
  e-prints, arXiv:1906.08009

\bibitem[{{Torres} {et~al.}(2019){Torres}, {Cantero}, {Rebassa-Mansergas},
  {Skorobogatov}, {Jim{\'e}nez-Esteban}, \& {Solano}}]{Torres2019}
{Torres}, S., {Cantero}, C., {Rebassa-Mansergas}, A., {et~al.} 2019, \mnras,
  485, 5573

\bibitem[{{Torres} {et~al.}(2001){Torres}, {Garc{\'{\i}}a-Berro}, {Burkert}, \&
  {Isern}}]{Torres2001}
{Torres}, S., {Garc{\'{\i}}a-Berro}, E., {Burkert}, A., \& {Isern}, J. 2001,
  \mnras, 328, 492

\bibitem[{{Wegg} {et~al.}(2015){Wegg}, {Gerhard}, \& {Portail}}]{Wegg2015}
{Wegg}, C., {Gerhard}, O., \& {Portail}, M. 2015, \mnras, 450, 4050

\end{thebibliography}

%-------------------------------------------------------------------

\end{document}